\documentclass[aps,pre,showpacs,twocolumn,superscriptaddress,longbibliography]{revtex4-1}
\usepackage[pdftex]{graphicx}
\usepackage[pdftex,bookmarks,colorlinks,breaklinks]{hyperref} % PDF hyperlinks, with coloured links
\hypersetup{linkcolor=red,citecolor=blue,filecolor=dullmagenta,urlcolor=blue} % colored links
\hypersetup{pdftoolbar=true,pdfpagemode=UseNone,pdfstartview=FitH, colorlinks=true,extension=}

\begin{document}
\title{Spectra and spectral correlations of microwave graphs with symplectic symmetry}
\author{A. Rehemanjiang}
\affiliation{Fachbereich Physik der Philipps-Universit\"at Marburg, D-35032 Marburg, Germany}
\author{M. Richter}
\affiliation{Universit\'{e} C\^{o}te d'Azur, CNRS, Institut de Physique de Nice (InPhyNi), 06108 Nice, France}
\author{U. Kuhl}
\affiliation{Fachbereich Physik der Philipps-Universit\"at Marburg, D-35032 Marburg, Germany}
\affiliation{Universit\'{e} C\^{o}te d'Azur, CNRS, Institut de Physique de Nice (InPhyNi), 06108 Nice, France}
\author{H.-J. St\"ockmann}
\affiliation{Fachbereich Physik der Philipps-Universit\"at Marburg, D-35032 Marburg, Germany}
\date{\today}
\begin{abstract}
Following an idea by Joyner et al.~[EPL, 107 (2014) 50004] a microwave graph with antiunitary symmetry $\mathcal{T}$
obeying $\mathcal{T}^2=-1$ has been realized. The Kramers doublets expected for such systems have been clearly
identified and could be lifted by a perturbation which breaks the antiunitary symmetry. The observed spectral level spacings distribution of the Kramers doublets is in agreement with the predictions from the Gaussian symplectic ensemble (GSE), expected for chaotic systems with such a symmetry. In addition results on the two-point correlation function, the spectral form factor, the number variance and the spectral rigidity are presented, as well as on the transition from GSE to GOE statistics by continuously changing $\mathcal{T}$ from $\mathcal{T}^2=-1$ to $\mathcal{T}^2=1$.
\end{abstract}


\maketitle

\section{Introduction}
Random matrix theory has proven to be an extremely powerful tool to describe the spectra of chaotic systems \cite{meh91,haa01b}.
For systems with time-reversal symmetry (TRS) and no spin 1/2 in particular there is an abundant number of
studies, both theoretical and experimentally, showing that their universal spectral properties are perfectly
well reproduced by the corresponding properties of the Gaussian orthogonal random matrix ensemble (GOE) (see
e.\,g.~Ref.~\onlinecite{stoe99} for a review). This is the essence of the famous conjecture by Bohigas,
Giannoni, Schmitt \cite{boh84b} (see also Ref.~\onlinecite{cas80}) which meanwhile has been proven in mayor parts by
the joint efforts of various groups \cite{ber85,sie01,muel09}. For systems with TRS and spin 1/2 the Gaussian
symplectic ensemble (GSE) holds instead, and for system without TRS the Gaussian unitary ensemble (GUE).
For the latter two of the classical ensembles the experimental situation is still unsatisfactory. There are
altogether only three studies of the spectra of systems with broken TRS showing GUE statistics \cite{so95,sto95b,hul04}, all
of them applying microwave techniques. For the GSE there is as yet only one recent experimental realization by our group
\cite{reh16}.
In the present paper a more detailed account on the latter work is given, as well as a number of new results.

This paper is organized as follows. In section~\ref{sec:theory} a theoretical description of graphs with
symplectic symmetry is given. Section~\ref{sec:experiment} describes our experimental realization of such
graphs. In section~\ref{sec:results} we present results on spectra, level spacings distribution and
two-point correlation functions. Furthermore, we discuss the change of the level
statistics when varying the antiunitary symmetry continuously from $\mathcal{T}^2=-1$ to $\mathcal{T}^2=1$.

\section{Theory}
\label{sec:theory}

\subsection{Graphs showing Kramer's degeneracy}
\label{sec:Kramers}

\begin{figure}
  \raisebox{4cm}[0cm][0cm]{(a)}\hspace*{0.5cm}\includegraphics[width=0.8\columnwidth]{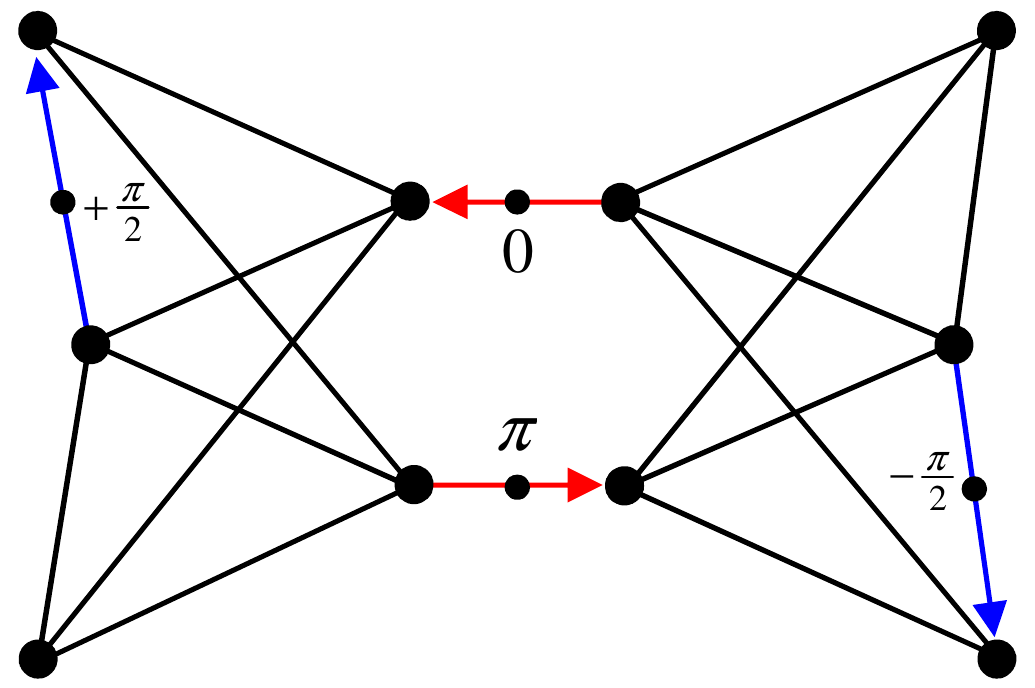}\hspace*{1.cm}\quad\\[1ex]
  \raisebox{4cm}[0cm][0cm]{(b)}\includegraphics[width=0.95\columnwidth]{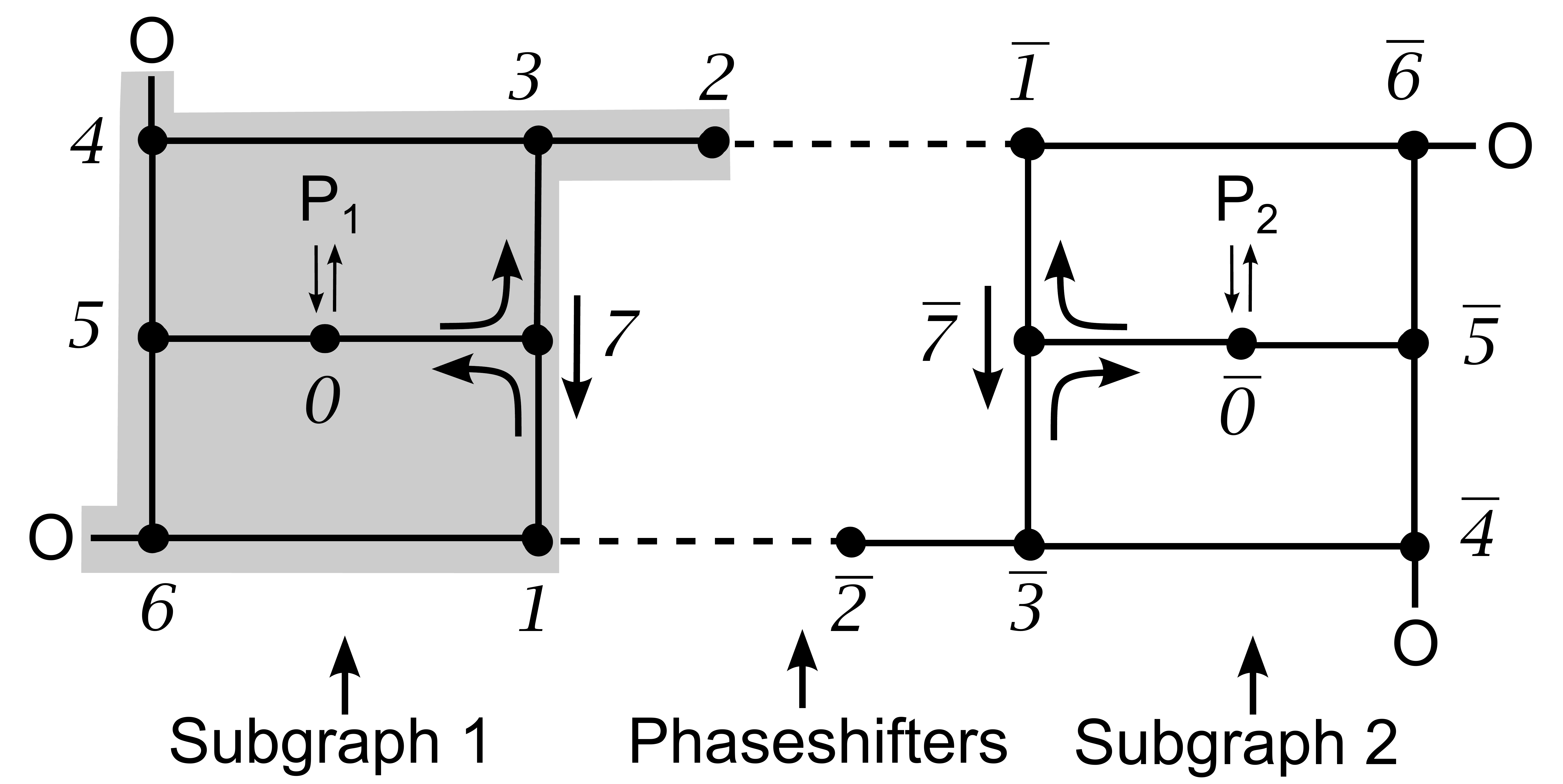}\\[1ex]
  \raisebox{7.5cm}[0cm][0cm]{(c)}\includegraphics[width=0.95\columnwidth]{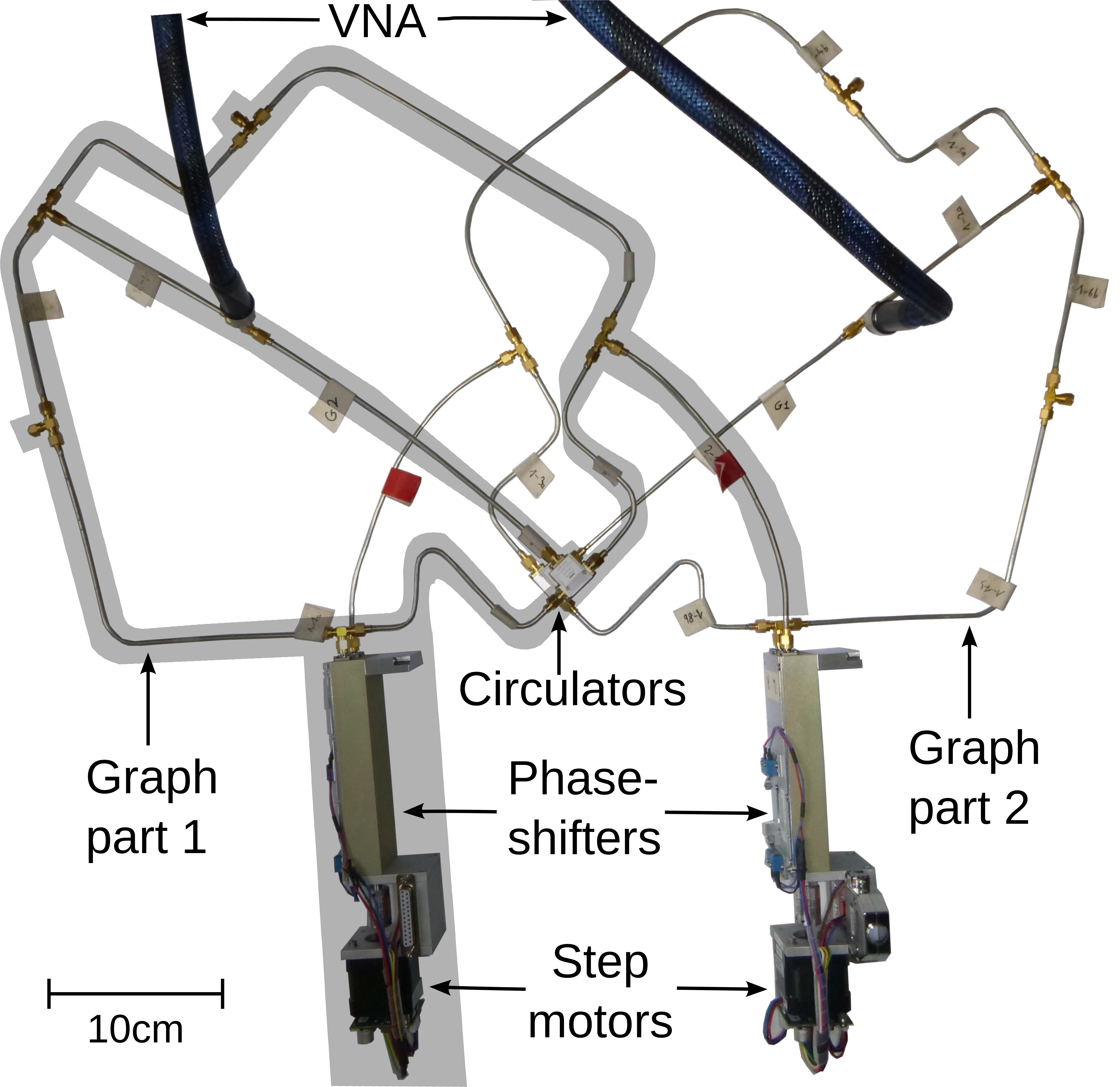}
  \caption{\label{fig:graph} (Color online)
    (a) Sketch of the graph proposed in Ref.~\onlinecite{joy14} to study GSE statistics without spin. The four arrows denote bonds along which additional phases are acquired.
	(b) Schematic drawing of one of the realized microwave graphs. Subgraph 1 is highlighted by a grey background.
    The dashed lines correspond to phaseshifters with variable lengths.
	The two subgraphs contain microwave circulators at nodes $7$ and $\bar{7}$, respectively, with opposite sense of rotation.
	The nodes marked by ``O'' are closed by open end terminators.
	They were used to allow for an easy realization of alternative graphs.
	Subgraphs 1 and 2 are connected at nodes $0$ and $\bar{0}$, respectively, to ports $P_1$ ($P_2$) of the VNA.
	(c) Photograph of the graph sketched in (b) consisting of T-junctions, semirigid cables with identification tags, circulators, open end terminators, and phaseshifters with step motors.
	Again subgraph 1 is highlighted.
	}
\end{figure}

Quantum graphs had been introduced by Kottos, Smilansky \cite{kot97a} as an ideal tool to study quantum chaos.
Just as for quantum billiards there is a one-to-one mapping onto the corresponding microwave graphs, where
the voltage at a node corresponds to the wave function and the current to the derivative of the wave function along the
bonds.
A comprehensive description of graphs can be found in the report by Kottos and Smilansky \cite{kot99a}.
Therefore, we restrict the discussion to the aspects relevant in the present context.

A quantum graph is constructed by a number of bonds, connected at vertices. In the present microwave realization the
bonds are coaxial cables connected by T-junctions. Hence, in our case the number of bonds meeting at a
vertex is always three. Each bond connecting two vertices $i$ and $j$ supports a one-dimensional wave function which may
be either written as
\begin{equation}\label{eq:psi1}
  \psi_{ij}(x_i)=a_{ij}e^{- \imath (k - A_{ij}) x_i}+b_{ij}e^{\imath (k + A_{ij}) x_i}\,,
\end{equation}
where $x_i$ is the distance from vertex $i$, or as
\begin{equation}\label{eq:psi2}
  \psi_{ji}(x_j)=a_{ji}e^{- \imath (k - A_{ji}) x_j}+b_{ji}e^{\imath (k + A_{ji}) x_j}
\end{equation}
where $x_j$ is the distance from vertex $j$. The $A_{ij}$ are vector potentials
resulting from a magnetic field if present. There are the relations
\begin{equation}\label{eq:xx}
  x_j= L_{ij}-x_i
\end{equation}
and
\begin{equation}\label{eq:ab}
  b_{ij}=a_{ij}e^{- \imath (k - A_{ij})L_{ij}}\,,\qquad b_{ji}=a_{ji}e^{- \imath (k - A_{ji}) L_{ij}}
\end{equation}
where $L_{ij}$ is the length of the bond connecting vertex $i$ to vertex $j$.
The scattering matrix $S$ relates the incoming amplitudes $a_{ij}$ with the outgoing ones $b_{ij}$ via
\begin{equation}\label{eq:scatt}
  b=Sa
\end{equation}
where $a$, $b$ are the vectors of incoming and outgoing amplitudes, respectively. For the T-junctions used
in the experiment all three ports are equivalent meaning a scattering matrix
symmetric with respect to a simultaneous change of rows and columns,
\begin{equation}\label{eq:Tscatt}
  S=\frac{1}{3}\left(\begin{array}{rrr}
    -1 & 2 & 2 \\
    2 & -1 & 2 \\
    2 & 2 & -1
  \end{array}\right)\,.
\end{equation}
There are two constraints to be met for the wave functions at the vertices:\\
(i) Continuity:
\begin{equation}
  \left.\psi_{ij}(x_i)\right|_{x_i=0}= \psi_i
\end{equation}
for all bonds connected to vertex $i$.\\
(ii) Current conservation:
\begin{equation}\label{eq:curr}
  \sum\limits_j\left.\left(-\imath A_{ij} + \frac{\partial}{\partial x_i}\right)\psi_{ij}(x_i)\right|_{x_i=0}=0
\end{equation}
where the sum is over all vertices $j$ connected to vertex~$i$.

Equation~(\ref{eq:curr}) holds for Neumann boundary conditions, the situation met in the experiment.
For microwave graphs these two conditions are equivalent to the well-known Kirchhoff relations governing electric circuits.
The continuity condition can be met by construction, just by choosing the $a_{ij}$ and the $b_{ij}$
appropriately. But current conservation implies a homogeneous linear equation system which has solutions only
if the determinant of the associated secular matrix $h(k)$ vanishes,
\begin{equation}\label{eq:det}
  \mathrm{det}[h(k)]=0
\end{equation}
where the matrix elements of $h(k)$ are given by
\begin{equation}\label{eq:sec}
  h_{ij}(k)=\left\{ \begin{array}{cc}
    -\sum\limits_{n\ne i}C_{in}\cot(kL_{im}) & i=i \\
    C_{ij}e^{-{\imath\varphi_{ij}}}\left[\sin(kL_{ij})\right]^{-1} & i\ne j
  \end{array}\right.\,.
\end{equation}
The $C_{ij}$ are the elements of the connectivity matrix, $C_{ij}=1$, if nodes $i$ and $j$ are connected, and
$C_{ij}=0$ otherwise. $\varphi_{ij}=A_{ij}L_{ij}$ is a phase resulting from a possible vector potential which, if present,
breaks TRS. The solutions of the determinant condition~(\ref{eq:det}) generate the spectrum of
the graph.

To realize graphs with GSE symmetry Joyner et\,al. \cite{joy14} proposed the graph shown in Fig.~\ref{fig:graph}.
It contains two geometrically identical subgraphs, but with phase shifts by $+\pi/2$ and $-\pi/2$, respectively, along two corresponding bonds.
The two subgraphs are connected by one pair of bonds yielding a graph with a geometric inversion center.
In addition there is another phase
shift by $\pi$ along one of the two bonds, but not the other one. This is the crucial point: Due to this
trick the total graph is symmetric with respect to an antiunitary operator $\mathcal{T}$, squaring to minus one,
$\mathcal{T}^2=-1$.

For the graph shown in Fig.~\ref{fig:graph} the secular matrix $h(k)$ may be written as
\begin{equation}\label{eq:sec1}
  h=h_{\mathrm{dis}}+v
\end{equation}
where $h_{\mathrm{dis}}$ is the secular matrix for the disconnected subgraphs, and $v$ describes the
connecting bonds. It is convenient to introduce an order of rows and columns according to $\{1, 2,
\dots,n;\bar{1}, \bar{2}, \dots, \bar{n}\}$, where the figures without bar refer to the vertices of subgraph
1, and the figures with bar to those of subgraph 2. $h_{\mathrm{dis}}$ may then be written as
\begin{equation}\label{eq:sec2}
  h_{\mathrm{dis}}=\left(
    \begin{array}{cc}
      h_0 & \cdot \\
      \cdot & h_0^* \\
    \end{array}
  \right)
\end{equation}
where $h_0$ and $h_0^*$ are the secular matrices for each of the two subgraphs, respectively. Since the only
difference between the subgraphs is the sign of the $\pi/2$ phase shift in one of the bonds, their secular
matrices are just complex conjugates of each other, see Eq.~(\ref{eq:sec}). Assuming for the sake of
simplicity that there is just one pair of bonds connecting node $1$ with node $\bar{2}$, and node $\bar{1}$
with node $2$, respectively, the matrix elements of $v$ are given by
\begin{eqnarray}
  v_{11}&=&v_{22}=v_{\bar{1}\bar{1}}=v_{\bar{2}\bar{2}}=-\cot (kl)\\\label{eq:sec3}
  v_{1\bar{2}}&=&v_{\bar{2}1}=-v_{2\bar{1}}=-v_{\bar{1}2}=\left[\sin (kl)\right]^{-1}\\
  v_{ij}&=&v_{\bar{i}\bar{j}}=v_{i\bar{j}}=v_{\bar{i}j}=0 \quad \mbox{otherwise}
\end{eqnarray}
where $l$ is the length of the bonds connecting $1$ with $\bar{2}$ and $\bar{1}$ with $2$. The generalization
to a larger number of bond pairs is straightforward.
Changing now the sequence of rows and columns to $\{1, \bar{1};2, \bar{2}; \dots; n, \bar{n}\}$, the
resulting $2n\times 2n$ matrix $\tilde{h}(k)$ may be written in terms of a $n\times n$ matrix with
quaternion matrix elements,
\begin{equation}\label{eq:sec4}
  [\tilde{h}(k)]_{nm}= \left[\mathrm{Re}(h_0)_{nm}+ v_{nm}\right]{\bf 1} -\mathrm{Im}(h_0)_{nm}{\bf\tau_z}-v_{n\bar{m}}{\bf\tau_y}
\end{equation}
where
\begin{equation}\label{eq:sec6}
    {\bf 1}=\left(
              \begin{array}{cc}
                1 & \cdot \\
                \cdot & 1 \\
              \end{array}
            \right),\quad
    {\bf \tau_z}=\left(
      \begin{array}{cc}
        -\imath & \cdot \\
        \cdot & \imath \\
      \end{array}
    \right),\quad
    {\bf \tau_y}=\left(
      \begin{array}{cc}
        \cdot & -1 \\
        1 & \cdot \\
      \end{array}
      \right)
\end{equation}
The determinant is not changed by this rearrangement of rows and columns,
$\mathrm{det}[h(k)]=\mathrm{det}[\tilde{h}(k)]$. The matrix elements $[\tilde{h}(k)]_{nm}$ commute with
$\mathcal{C}{\bf\tau_y}$, where $\mathcal{C}$ denotes the complex conjugate, and hence the whole matrix commutes with
\begin{equation}\label{T}
  \mathcal{T}=\mathrm{diag}(\mathcal{C}{\bf\tau_y}\, \dots,\mathcal{C}{\bf\tau_y})\,,
\end{equation}
where $\mathcal{T}$ squares to minus one,
\begin{equation}\label{t2}
  \mathcal{T}^2=-1\,.
\end{equation}
This is exactly the situation found for spin 1/2 systems, as in such systems a two-fold Kramers
degenerate spectrum is expected showing the signatures of the GSE provided the system is chaotic, see e.\,g.\
Chapter~2 of Ref.~\onlinecite{haa01b}~. The two essential ingredients had been two subgraphs with secular
matrices $h_0(k)$ and $h_0^*(k)$, being complex conjugates of each others, see Eq.~(\ref{eq:sec2}), and the additional
phase shift of $\pi$ applied in one of the connecting bonds giving rise to the minus signs in Eq.~(\ref{eq:sec3}).

\subsection{Scattering properties of symplectic graphs}
\label{sec:scatt}

For a measurement of the spectral properties of a graph cables have to be attached to it. To maintain the
antiunitary symmetry, cables must come in pairs, attached at symmetry equivalent points, e.\,g. at the upper
left and the lower right vertex, respectively of the graph shown in Fig.~\ref{fig:graph}. Again we can rely
on the results of Kottos, Smilansky \cite{kot99a}, who derived an expression for the scattering matrix
describing this situation
\begin{equation}\label{eq:scat1}
  S=2iW^T\left[h(k)+\imath WW^T\right]W-1
\end{equation}
(with a small change in notation compared to Ref.~\onlinecite{kot99a}), where $h(k)$ is the matrix defined in
Eq.~(\ref{eq:sec1}), and $W$ is a $N\times L$ matrix describing the coupling to the environment, where
$N$ is the number of vertices, and $L$ the number of channels.
The scattering matrix $S$ is thus a $L\times L$ matrix.
For the microwave system each attached cable corresponds to one of these channels. In microwave
graphs the determination of the coupling constants is straightforward. All microwave components obey the 50\,$\Omega$
convention, meaning that a cable connecting the network analyzer to the graph corresponds to an ideally
matched open channel. For the matrix elements of $W$ this means that $W_{vl}=1$ if channel $l$ is attached
to vertex $v$, and $W_{vl}=0$ for all other cases.
The scattering matrix~(\ref{eq:scat1}) is unitary by construction. Experimentally obtained scattering matrices, however,
are always sub-unitary because of absorption. There are standard techniques to take care of absorption in terms of a large
number of weakly coupled fictitious channels (see e.\,g.\ Ref.~\onlinecite{sav06}). To keep the discussion simple, this aspect will be postponed to section~\ref{sec:experiment}.

Equation~(\ref{eq:scat1}) may be elementarily transformed into
\begin{equation}\label{eq:scat2}
  S=-\frac{1-iW^TGW}{1+iW^TGW}\,, \qquad \mbox{where\hspace{2em}} G=h^{-1}\,.
\end{equation}
Up to the sign this is the expression for the scattering matrix familiar in the context of quantum billiards
and quantum dots, where
\begin{equation}\label{eq:defG}
  G=\frac{1}{E-H}
\end{equation}
and $H$ is the Hamiltonian of the system. The minus sign is a consequence of the definitions of the
$a_{ij}$ and $b_{ij}$ in Eqs.~(\ref{eq:psi1}), (\ref{eq:psi2}), and (\ref{eq:Tscatt}) which have been chosen
in accordance with Ref.~\onlinecite{kot99a}. In the context of quantum billiards and quantum dots, however,
usually another convention is applied, with a minus sign for the $b_{ij}$ in Eqs.~(\ref{eq:psi1}) and
(\ref{eq:psi2}). As a consequence the resulting scattering matrices differ in sign
(compare e.\,g.\ Eq.~(23) in Ref.~\onlinecite{kot99a} and Eq.~(81) in Ref.~\onlinecite{bee97}).

Since $h$ commutes with $\mathcal{C}{\bf\tau_y}$, and $W$ is real, application of $\mathcal{T}$ to $S$ yields
\begin{eqnarray}\label{eq:TScalc}\nonumber
  \mathcal{T}S&=&\mathcal{T}\left(-\frac{1-iW^TGW}{1+iW^TGW}\right) \\
  &=& \left(-\frac{1+iW^TGW}{1-iW^TGW}\right)\mathcal{T}=S^{-1}\mathcal{T}
\end{eqnarray}
or
\begin{equation}\label{eq:TS}
  S=\mathcal{T}^{-1}S^{-1}\mathcal{T}=\mathcal{T}^{-1}S^\dag \mathcal{T}\,.
\end{equation}
If there are only two channels, as in the present experiment, $S$ is a $2\times 2$ matrix
\begin{equation}\label{eq:S}
  S=\left(
    \begin{array}{cc}
      a & b \\
      c & d \\
    \end{array}
  \right)\,.
\end{equation}
For this special situation Eq.~(\ref{eq:TS}) yields
\begin{equation}\label{eq:TST}
    S=\mathcal{T}^{-1}S^\dag \mathcal{T}=-\tau_y \mathcal{C} \left(
        \begin{array}{cc}
          a^* & c^* \\
          b^* & d^* \\
        \end{array}
      \right) \mathcal{C}\tau_y=
\left(
        \begin{array}{rr}
          d & -b \\
          -c & a \\
        \end{array}
      \right)
\end{equation}
whence follows $a=d$ and $b=c=0$, or
\begin{equation}
    S=e^{\imath \alpha}\left(
                   \begin{array}{cc}
                     1 & \cdot \\
                     \cdot & 1 \\
                   \end{array}
                 \right)\,.
\end{equation}
The antiunitary symmetry $\mathcal{T}$ with $\mathcal{T}^2=-1$ thus implies that there is no transmission!
The information on the graph properties is thus encoded in the reflection phase $\alpha$. Denoting both the
vertex, where of the cable is coupled to the graph, and the channel corresponding to the cable by `0', the
corresponding $S$ matrix
element $S_{00}$ is given by
\begin{equation}\label{eq:s00}
  S_{00}= e^{\imath \alpha}=-\frac{1-iG_{00}}{1+iG_{00}}
\end{equation}
where we have used
\begin{equation}
  \left(W^T G W\right)_{00}=\sum\limits_{kl}W_{k0}G_{kl}W_{l0}=G_{00}\,.
\end{equation}
It follows for the phase
\begin{equation}\label{eq:alpha}
  \alpha=\pi -2\arctan G_{00}\,.
\end{equation}
$G_{00}=\left[h(k)^{-1}\right]_{00}$ may be expanded into partial fractions,
\begin{equation}\label{eq:G00}
  G_{00}=\sum\limits_n\frac{a_n}{k-k_n}
\end{equation}
where the $k_n$ are the zeros of the secular determinant $\left|h(k)\right|$. Due to Eq.~(\ref{eq:alpha})
these zeros show up in steps of the reflection phase. We shall see that this allows for a very convenient
extraction of the eigenvalues $k_n$.

\subsection{The single pair of bonds approximation}\label{sec:twobond}

The peculiar symmetry of the studied graph ensures that all eigenvalues are two-fold degenerate. But this does not
automatically imply that a single pair of bonds is already sufficient to turn the two identical GUE spectra of the
disconnected subgraphs into one GSE spectrum for the total graph. For a study of this question the secular
matrix~(\ref{eq:sec}) is not very convenient. Therefore, we now turn to a description of the graph in terms of
a Hamiltonian. Scattering theory yields for the total Hamiltonian $H$ of two coupled subgraphs
\begin{equation}\label{eq:ham}
  H=\left(
    \begin{array}{cc}
      H_0  & V\bar{V}^\dag \\
      \bar{V}V^\dag & \bar{H}_0 \\
    \end{array}
  \right)
\end{equation}
Here $H_0$ and $\bar{H}_0$ are the Hamiltonians of the two disconnected subgraphs, and the off-diagonal blocks
describe the coupling. In the eigenbases of $H_0$ and $\bar{H}_0$ the diagonal blocks are diagonal with
matrix elements
\begin{equation}\label{eq:ham0}
  \left(H_0\right)=E_n^0\delta_{nm}\,, \qquad \left(\bar{H}_0\right)=\bar{E}_n^0\delta_{nm}
\end{equation}
where the $E_n^0$ and $\bar{E}_n$ are the eigenvalues of the disconnected graphs with Neumann boundary
conditions at the coupling point. $V$ and $\bar{V}$ are $N\times K$ matrices, where $N$ is the rank of the
Hamiltonians $H_0$ and $\bar{H}_0$, assumed to be the same for the sake of simplicity, and $K$ is the
number of coupling bonds. In the eigenbases of $H_0$ and $\bar{H}_0$ the elements of $V$ are just the values
of the wave functions of the disconnected subgraphs at the coupling points, $V_{nk}=\psi_n(x_k)$ and
$\bar{V}_{nk}=\bar{\psi}_n(x_k)$, respectively.

In the present situation $H$ and $\bar{H}$ are complex conjugates of each other, $\bar{H}=H^*$,
whence follows $\bar{E}_n=E_n^0$ and $\bar{V}_{nk}=V_{nk}^*$. Let us further assume that there is just a single
pair of bonds coupling node $1$ with node $\bar{2}$ and node $2$ with node $\bar{1}$, respectively,
one with a phase shift 0, the other one with a phase shift $\pi$ at the coupling point. The matrix elements
of the off-diagonal block are then given by
\begin{eqnarray}\label{eq:ham1}
  \left[V\bar{V{^\dag}}\right]_{nm}&=&V_{n1}\bar{V}_{m\bar{2}} + e^{\imath \pi} V_{n2}\bar{V}_{m\bar{1}}\\
  \nonumber&=&\psi_{n1}\psi_{m2}-\psi_{n2}\psi_{m1}=\tilde{V}_{nm}
\end{eqnarray}
and $ [\bar{V}V{^\dag}]_{mn}=-\tilde{V}_{nm}^*$, with the abbreviation $\psi_{nk}=\psi_n(x_k)$.
In the discussion of the spacing distribution of neighbored levels in chaotic systems the Wigner
approximations have proven extremely successful. They are exact for $2\times 2$ Gaussian random matrix
ensembles, and deviate from the exact results for large rank matrices only by several percents (see e.\,g.\
Ref.~\onlinecite{haa01b}) too small to be observed in most experimental spectra. Therefore, usually the Wigner
approximation has been used for the interpretation of the spectra. We shall follow this practice in the present
work. Exactly in the spirit of the Wigner approximation we restrict the rank of the Hamiltonian $H$ in Eq.~(\ref{eq:ham}) to 2.
Without loss of generality we may take the average energy as 0,
\begin{equation}
  E_{1/2}=\pm\frac{a}{2}\,.
\end{equation}
The characteristic polynomial of the Hamiltonian is now given by
\begin{equation}\label{eq:chi}
  \chi(E)=\left|
    \begin{array}{cccc}
      E-a/2  & \cdot & \cdot & -\tilde{V}_{12} \\
      \cdot & E+a/2 & \tilde{V}_{12} & \cdot \\
      \cdot & \tilde{V}_{12}^* & E-a/2 &  \cdot \\
      -\tilde{V}_{12}^* & \cdot &\cdot & E+a/2   \\
    \end{array}
  \right|\,.
\end{equation}
$\chi(E)$ factorizes in to a product of two identical quadratic equations,
\begin{equation}\label{eq:chi1}
  \chi(E)=\left[E^2-(a/2)^2-\left|\tilde{V}_{12}\right|^2\right]^2\,.
\end{equation}
Each eigenvalue is thus two-fold degenerate, again a manifestation of Kramers degeneracy. The
distance of the two eigenvalues is given by
\begin{equation}\label{eq:dist}
  s=\sqrt{a^2+4\left|\tilde{V}_{12}\right|^2}
\end{equation}
whence follows for the level spacing distribution function,
\begin{eqnarray}\label{eq:ps}
  p_0(s)&=&\left<\delta\left(s-\sqrt{a^2+4\left|\tilde{V}_{12}\right|^2}\right)\right>\\\nonumber
  &=&\int \delta\left(s-\sqrt{a^2+z}\right)p_a(a)p_V(z)\,\mathrm{d}a\,\mathrm{d}z
\end{eqnarray}
where $p_a(a)$ and $p_V(z)$ are the distribution functions of $a$ and $4|\tilde{V}_{12}|^2$, respectively.
Since the circulators break time-reversal symmetry in the subgraphs \cite{hul04}, see below, we expect a Wigner
GUE level spacing distribution for each subgraph,
\begin{equation}\label{eq:pa}
  p_a(a)= p_{\,\mathrm{GUE}}(a)
\end{equation}
where
\begin{equation}\label{eq:GUE}
  p_{\,\mathrm{GUE}}(s)=\frac{32}{\pi^2}s^2\exp\left(-\frac{4}{\pi}s^2\right)
\end{equation}
is the Wigner level spacing distribution for the GUE.
For the calculation of $p_V(z)$ we assume that real and imaginary parts $\psi_R$, $\psi_I$ of all wave
function entering expression~(\ref{eq:ham1}) for $\tilde{V}_{12}$ are uncorrelated and Gaussian distributed,
\begin{equation}\label{eq:gauss}
  p_\psi\left(\psi_R,\psi_I\right)=\frac{1}{\pi\langle\left|\psi\right|^2\rangle}\exp\left(-\frac{\left|\psi\right|^2}{\langle\left|\psi\right|^2\rangle}\right)
\end{equation}
From the normalization $\int_L |\psi|^2 \mathrm{d}l=1$, where the integral is over all bonds, it follows
\begin{equation}\label{eq:normL}
  \langle\left|\psi\right|^2\rangle=\frac{1}{L}
\end{equation}
where $L$ is the total length of all bonds.

\begin{figure}
  \raisebox{4cm}[0cm][0cm]{}\includegraphics[width=0.95\columnwidth]{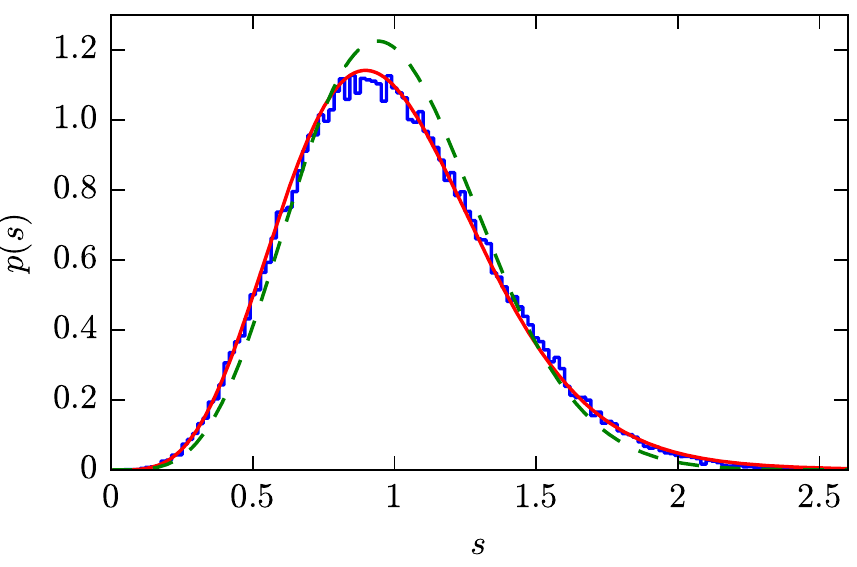}\\[1ex]
  \caption{\label{fig:ps} (color online)
  Level spacing distribution for the single pair of bonds approximation (solid line), see Eq.~(\ref{eq:psres}), and for the Wigner GSE distribution (dashed line), see Eq.~(\ref{eq:GSE}). In addition the result of a random matrix simulation is shown (blue histogram). In the simulations we created random matrices for an ensemble of 2000 matrices of size $1000 \times 1000$, i.e.~the size of each GUE subblock $H_0$ and $\bar{H}_0$, see Eq.~(\ref{eq:ham}), was $500 \times 500$. We used only the central 0.1 fraction of each spectrum to calculate the distribution of next-nearest neighbor spacings $s$ in order to avoid problems with the non-constant density of states.
  }
\end{figure}

For a graph being a one-dimensional system the mean density of states is constant if the spectrum is taken as a function of $k$ and is given by Weyl's law \cite{kot99a}
\begin{equation}\label{eq:weyl}
  \rho_\mathrm{Weyl}(k)= L/\pi\,.
\end{equation}
Since Eq.~(\ref{eq:GUE}) assumes a constant mean level spacing of one, we have to consider the spectrum on a $k$ axis and have to scale the total length to
$L=\pi$ to obtain a mean level spacing of $\bar{\rho}=1$.
Equation~(\ref{eq:gauss}) then reads
\begin{equation}\label{eq:gauss1}
  p_\psi\left(\psi_R,\psi_I\right)=e^{-\pi|\psi|^2}\,.
\end{equation}
With this ingredient $p_V(z)$ can be calculated yielding
\begin{equation}\label{eq:pv}
  p_V(z)=\frac{\pi^3}{4}\sqrt{z}K_1\left(\pi\sqrt{z}\right)
\end{equation}
where $K_1(t)$ is a modified Bessel function. The technical details are given in the appendix.
Entering with expressions~(\ref{eq:pa}) and (\ref{eq:pv}) for $p_a(a)$ and $p_V(z)$, respectively, into Eq.~(\ref{eq:ps}),
one obtains after substituting $a=r\cos\varphi$, $\sqrt{z}=r\sin\varphi$,
\begin{equation}\label{eq:p0s}
  p_0(s)=16s^4\!\int\limits_0^\pi \mathrm{d}\varphi \sin\varphi\cos^2\!\varphi\,e^{-\frac{4}{\pi}s^2\cos^2\!\varphi}\hat{K}_1(\pi
s\sin\varphi)
\end{equation}
where we have introduced $\hat{K}(t)=tK(t)$. The factor $t$ compensates the singularity of $K(t)$ for $t\to 0$.
$\hat{K}(t)$ is
regular for $t\to 0$, $\hat{K}(0)=1$. $p_0(s)$ thus shows a quartic level repulsion for $s\to0$, just as the
Wigner GSE distribution.
$p_0(s)$ is normalized by construction,
\begin{equation}\label{eq:norm_p0s}
  \int\limits_0^\infty p_0(s) \mathrm{d}s=1
\end{equation}
but this is not true for the mean level spacing,
\begin{equation}\label{eq:norm_bars}
  \bar{s}= \int\limits_0^\infty sp_0(s) \mathrm{d}s=1.32417\dots
\end{equation}
Therefore, in the last step $p_0(s)$ is scaled to the new level spacing distribution
\begin{equation}\label{eq:psres}
  p(s)= \bar{s}p_0(\bar{s}s)
\end{equation}
showing a mean level spacing of one,
\begin{equation}\label{eq:norm_ps}
  \langle s\rangle= \int\limits_0^\infty sp(s) \mathrm{d}s=1\,.
\end{equation}
In Fig.~\ref{fig:ps} the single pair of bonds approximation is compared to the Wigner GSE prediction,
\begin{equation}\label{eq:GSE}
  p_{\,\mathrm{GSE}}(s)=\frac{2^{18}}{3^6\pi^3}s^4\exp\left(-\frac{64}{9\pi}s^2\right)\,.
\end{equation}
 Both distributions agree which each other up to relative deviations of at most some percent. Additionally, we performed random matrix simulations for the single pair of bonds approximation with matrices of rank $N=1000$. The resulting level spacing distribution nicely follows the theoretical prediction. This shows that the two-level approximation for the Hamiltonian $H$, being exact for $N=4$ matrices, is working well also in the large $N$ limit, just as it is the case for the Wigner distributions.

\section{Experiment}\label{sec:experiment}

\subsection{The set-up}

The requirements defined by Joyner et\,al. \cite{joy14} to realize graphs with GSE symmetry pose some
challenges. Since we did not know of a simple way to achieve phase shifts of $\pm\pi/2$ along the bonds, we
proceeded in a somewhat different way and instead built two geometrically identically subgraphs, but with two
circulators of opposite sense of rotation within the two subgraphs. A circulator is a microwave device
introducing directionality: Microwaves entering through ports 1, 2, 3 exit via ports 2, 3, 1, respectively.
The result is the same as with the $\pm\pi/2$ shifts:
the circulators break TRS, resulting in identical GUE spectra for the two subgraphs, but
with an opposite sense of propagation within the respective subgraphs. Again the two subgraphs may thus be
described in terms of a secular matrix $h_0$ and its complex conjugate $h_0^*$.

In the first step we restricted ourselves to only one pair of bonds between the two subgraphs. In order to achieve a
phase jump of $\pi$ along one of the bonds we tried different options. First we inserted pairs of circulators in each
bond oriented such that the waves could pass in both directions. Both open ends were closed with short-end terminators
in one bond and open-end terminators in the other one. This should result in the wanted phase difference of $\pi$
between the two bonds. The idea worked in principle, but the phase difference showed up not to be $\pi$ but be somewhat
larger, about $1.1\,\pi$. Probably this could be improved by a suitable fine-tuning of the terminators, but for the
moment we discarded this option. Next we tried it with I\&Q vector modulators (Model: M2L-68N-S from GT Microwave Inc.) which allow to adjust arbitrary phases and amplitudes in transmission. The observed spectra, however, showed up to be mostly independent on the phase difference imposed. We attribute this to the large insertion loss of 12 dB of the IQ modulators.
We finally solved the problem by means of mechanical phase shifters changing the phase within the bonds by a change of
its length (just like in a trombone). This approach has the obvious shortcoming that for a given optical length change $\Delta l$ the phase shift $\Delta\varphi$ depends on frequency $\nu$:
\begin{equation}\label{eq:phase}
  \Delta\varphi= k\Delta l= \frac{2\pi\nu}{c}\Delta l
\end{equation}
where $k$ is the wave number, and $c$ is the vacuum velocity of light.

Figure~\ref{fig:graph}(b) shows a schematic drawing and Fig.~\ref{fig:graph}(c) a photograph of one of the graphs used in the experiment.
The bonds of the graphs were formed by Huber~\&~Suhner EZ-141 coaxial semi-rigid cables with SMA connectors, coupled by T-junctions at the vertices.
The phase shifters (ATM, P1507) had been equipped with motors to allow for an automatic stepping. Reflection
and transmission measurements were performed with an Agilent 8720ES vector network analyzer (VNA) with two
ports at equivalent positions of the two subgraphs. The corresponding reflection and transmission amplitudes
will be denoted in the following by $S_{ij}$, $i,j=0,\bar{0}$. The operating range of the circulators (Aerotex
I70-1FFF) extended from 6 to 12\,GHz. Therefore, the evaluation of the spectra was restricted to this
window.

\subsection{Transmission measurements}

\begin{figure}
  \mbox{\raisebox{5.0cm}[0cm][0cm]{a)\hspace{-.5cm}}\includegraphics[width=\columnwidth]{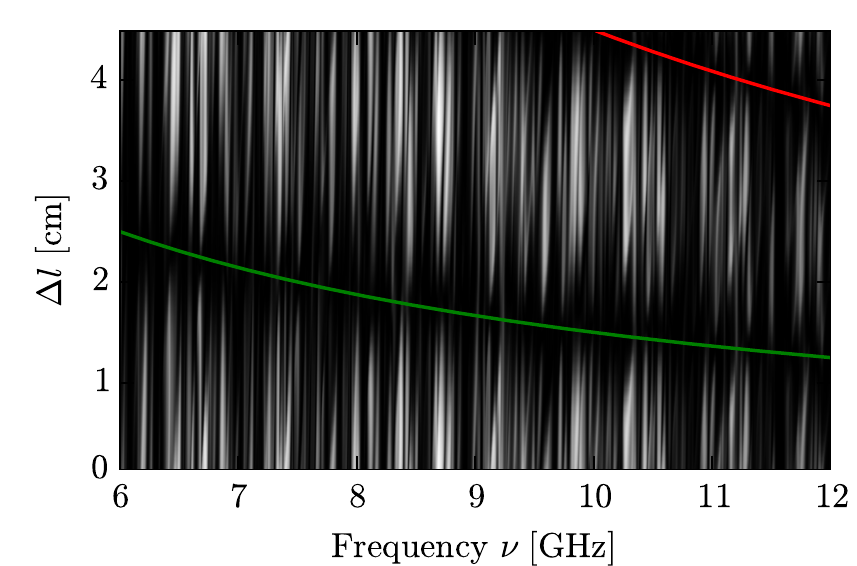}}\\
  \mbox{\raisebox{5.0cm}[0cm][0cm]{b)\hspace{-.5cm}}\includegraphics[width=\columnwidth]{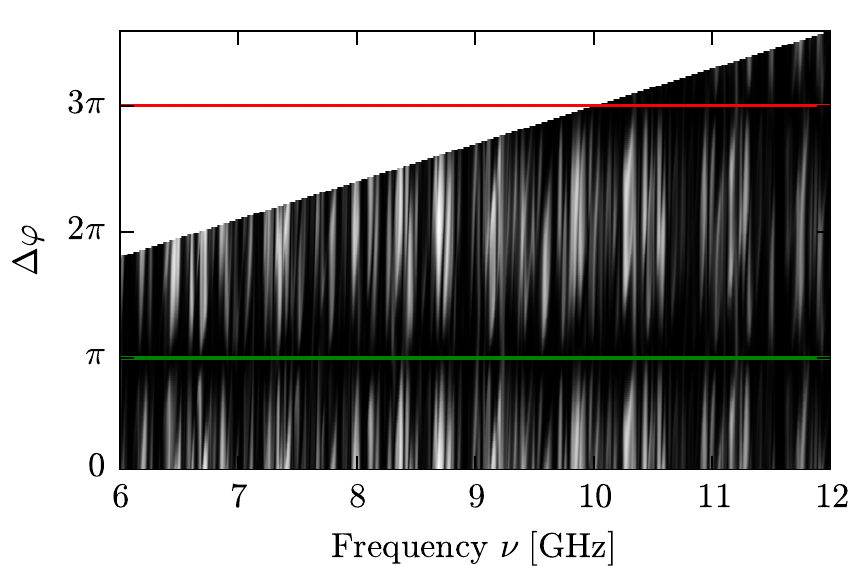}}
  \caption{\label{fig:S12_spectra}  (color online)
  (a) Transmission $\left|S_{0\bar{0}}\right|^2$ in dependence of frequency in a gray scale. Spectra for different $\Delta l$ are stacked onto each other.
  (b) The same data, but rearranged to yield spectra for constant $\Delta\varphi$.
  The green (red) solid line corresponds to constant $\Delta\varphi=\pi$ ($3\pi$), i.e.~to Eq.~\ref{eq:phase}.
  }
\end{figure}

We started by taking a series of measurements for constant $\Delta l$. Figure~\ref{fig:S12_spectra}(a) shows the
transmission for altogether 396 $\Delta l$ values stacked onto each other between $\Delta l_{\mathrm{min}}\approx 0$ and
$\Delta l_{\mathrm{max}}= 4.4$\,cm in a grey scale. The lines for $\Delta\varphi=\pi$ and $\Delta\varphi=3\pi$ are marked
in red and green, respectively. Next, a variable transformation from $\Delta l$ to $\Delta\varphi$ was performed, using
Eq.~(\ref{eq:phase}), to obtain the transmission $S_{0\bar{0}}$ for constant $\Delta\varphi$. The result is shown in
Fig.~\ref{fig:S12_spectra}(b).
For a given frequency $\nu$ the maximum $\Delta\varphi$ accessible is, according to Eq.~(\ref{eq:phase}), given by
$\Delta\varphi_{\mathrm{max}}=(2\pi\Delta l_{\mathrm{max}}/c)\nu$. The inaccessible regime above this limit is left white
in Fig.~\ref{fig:S12_spectra}(b). As expected the pattern is periodic in $\Delta\varphi$ with period $2\pi$. For
$\Delta\varphi=\pi$ and $\Delta\varphi=3\pi$ the transmission is strongly suppressed.
This is exactly what is expected: for GSE graphs the total transmission between equivalent nodes $i$ and $\bar{i}$ is
zero, see section~\ref{sec:scatt}.
Alternatively this may be explained in terms of an interference effect: All transmission paths from $P_1$ to $P_2$ come in
pairs.
One of them passes through one phase shifter whereas its partner passes through the other one, and as a result their
lengths differ by $\Delta l$. Depending on the resulting $\Delta\varphi$ this gives rise to constructive or destructive
interference.

\begin{figure}
  \includegraphics[width=\columnwidth]{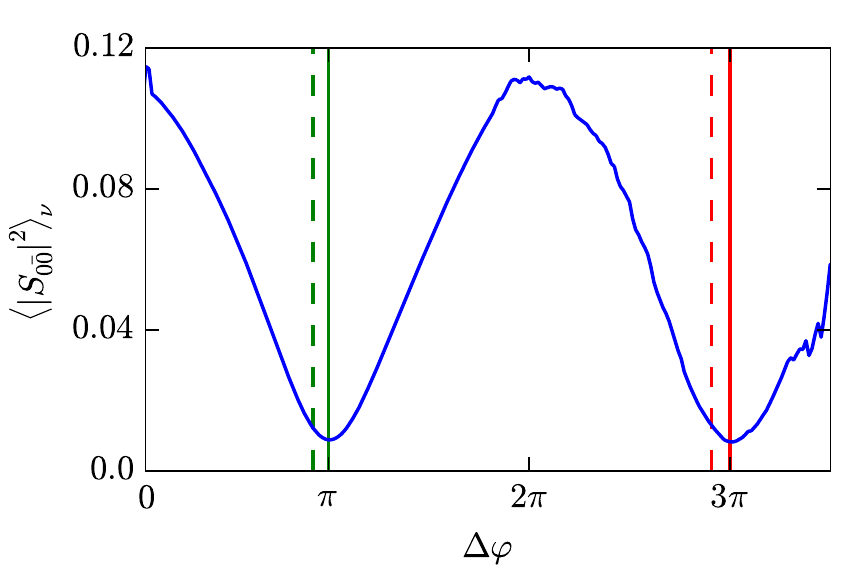}
  \caption{\label{fig:projection} (color online)
  Transmission $\langle \left|S_{0\bar{0}}\right|^2\rangle_\nu$, averaged over all frequencies $\nu$ between 6 and 12\,GHz in dependence of $\Delta\varphi$.
  See the text for the discussion of the solid and dotted lines.}
\end{figure}

In Fig.~\ref{fig:projection} the averaged transmission over all frequencies between 6 and 12\,GHz is plotted in dependence
of $\Delta\varphi$ meaning a projection of all data onto the $\Delta\varphi$ axis. This results in a considerable visual
enhancement of the interference effect. The transmission, which should be zero in the minima, still amounts to about 10\,\% of
the maximal transmission. This is due to the unavoidable tolerances in the construction of the two subgraphs.
It was one of our main concerns when we started the experiment, whether it would be possible to construct two
sufficiently equal subgraphs to see the wanted effects. Figure~\ref{fig:projection} illustrates that these doubts were unjustified.
The solid red and green lines mark the positions of the minima. They have been associated with $\Delta\varphi=\pi$ and
$\Delta\varphi=3\pi$. This corresponds to a calibration of the length difference $\Delta l$ via the experiment. In
addition there are dotted lines. These positions have been obtained by calculating $\Delta l$ from the known lengths of
the phase shifters and the cables in the bonds. The mismatch between the expected and observed positions of the minima
corresponds to an optical length of about 1.4\,mm. In view of the overall optical length of typically 2.8\,m of the used graphs
the mismatch in the expected and observed position of the minima is clearly within the limits of tolerance guaranteed by
our workshop. We thus felt justified to readjust $\Delta l$ using the transmission minima.

\section{Results}\label{sec:results}

\subsection{Spectra}\label{sec:spectra}

\begin{figure}
  \includegraphics[width=\columnwidth]{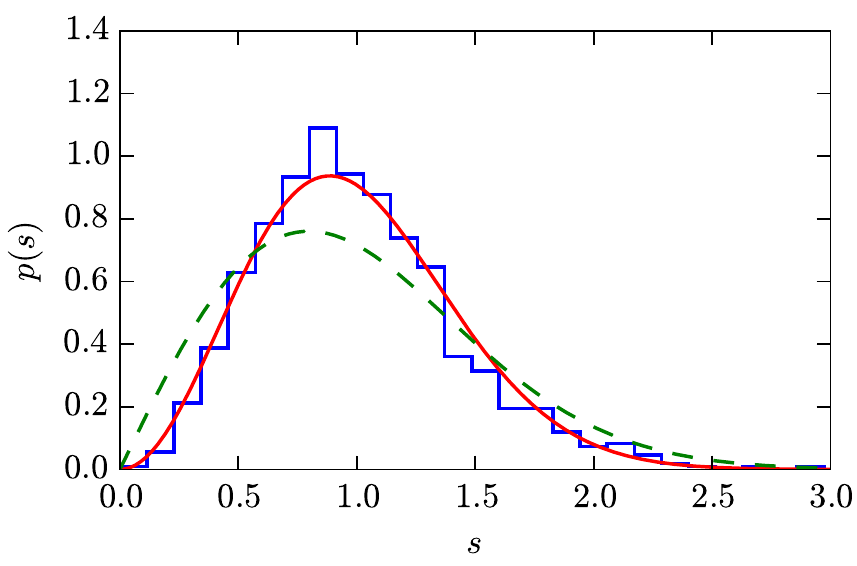}
  \caption{\label{fig:gue} (color online)
  Level spacing distribution for the disconnected subgraph. The results of 7 different subgraphs have been superimposed to improve the statistics. The solid line corresponds to the Wigner GUE distribution~(\ref{eq:GUE}), the dashed one to the GOE distribution.
  }
\end{figure}

\begin{figure}
  \includegraphics[width=\columnwidth]{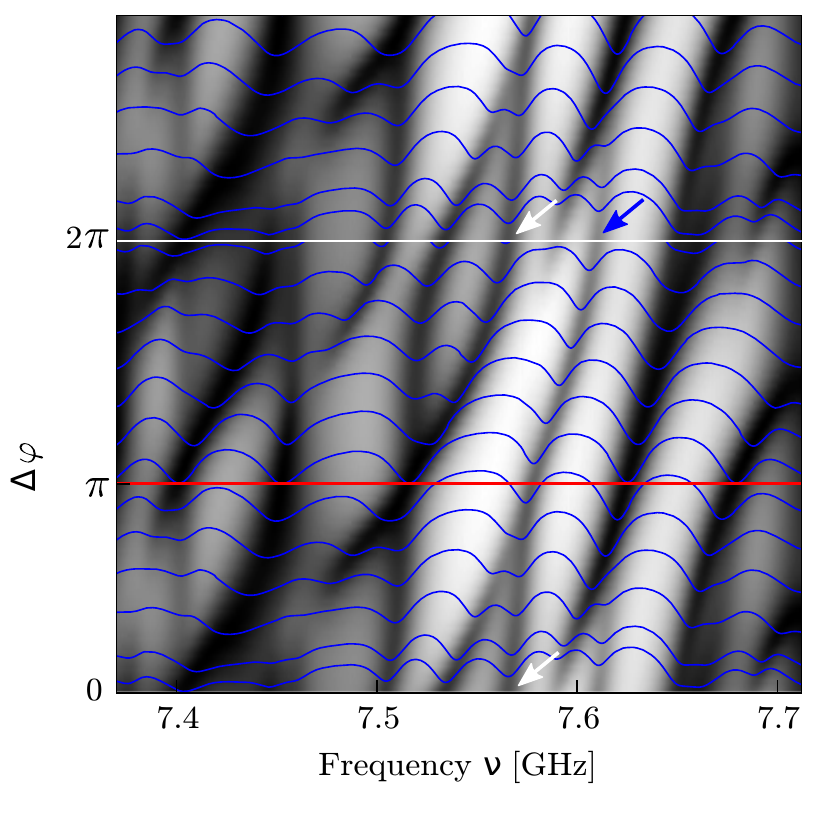}
  \caption{\label{fig:S11_spectra} (color online)
  Reflection $\left|S_{00}\right|^2$ in dependence of frequency in a gray scale. The spectra for different $\Delta \varphi$ are stacked onto each other. The bottom part of the spectra has been repeated at the top to emphasize the periodic structure. For the colored arrows see the discussion in section~\ref{sec:GSE_GOE}.}
\end{figure}

The circulators in the two subgraphs break time-reversal symmetry, and hence we expect GUE statistics for the two subgraphs
\cite{hul04}. To check this we started with a measurement of the reflection spectra for both disconnected subgraphs and found
them identical within the limits of resolution. Figure~\ref{fig:gue} shows the corresponding level spacing distribution
being in good agreement with the expected Wigner GUE distribution~(\ref{eq:GUE}).

Next we measured the spectra of the combined graph with a singe pair of connecting bonds. Because of the lack of transmission at $\Delta\varphi=\pi$ we instead analyzed the reflection $\left|S_{00}\right|^2$.
The results are shown in Fig.~\ref{fig:S11_spectra} for a small frequency window and for different $\Delta\varphi$, again stacked on top of each other in a shaded plot. The bottom part of the spectra has been repeated at the top to accentuate the periodicity of the spectra as a function of $\Delta\varphi$ with with a period of $2\pi$. Each eigenfrequency shows up as a dip. One clearly observes the formation of Kramers doublets at the $\pi$ line, and their splitting into singlets when departing from this line. There is a complete equivalence to the Zeeman splitting of spin doublets: In the present experiment the antiunitary symmetry is destroyed when departing from the $\pi$ line, whereas for conventional spin systems this effect occurs when applying a magnetic field. This is a clear confirmation that we have been successful in constructing a graph with anti-unitary symmetry $\mathcal{T}$ obeying $\mathcal{T}^2=-1$. The mutual distances between the six Kramers doublets seen in Fig.~\ref{fig:S11_spectra} at the $\pi$ line are the same within 20 percent. This shows a clear tendency of the levels towards an equal level spacings at the $\pi$ line, one of the fingerprints of a GSE spectrum.

Careful readers may wonder why there are visible resonances at all in the reflection, since, if there is no
transmission, the reflection should be one everywhere. The explanation for this apparent contradiction is absorption
within the bonds of the graphs, which had not been considered in the derivation of section~\ref{sec:scatt}. It leads to a dip in the reflection whenever an eigenfrequency is excited.
\begin{figure}
  \includegraphics[width=\columnwidth]{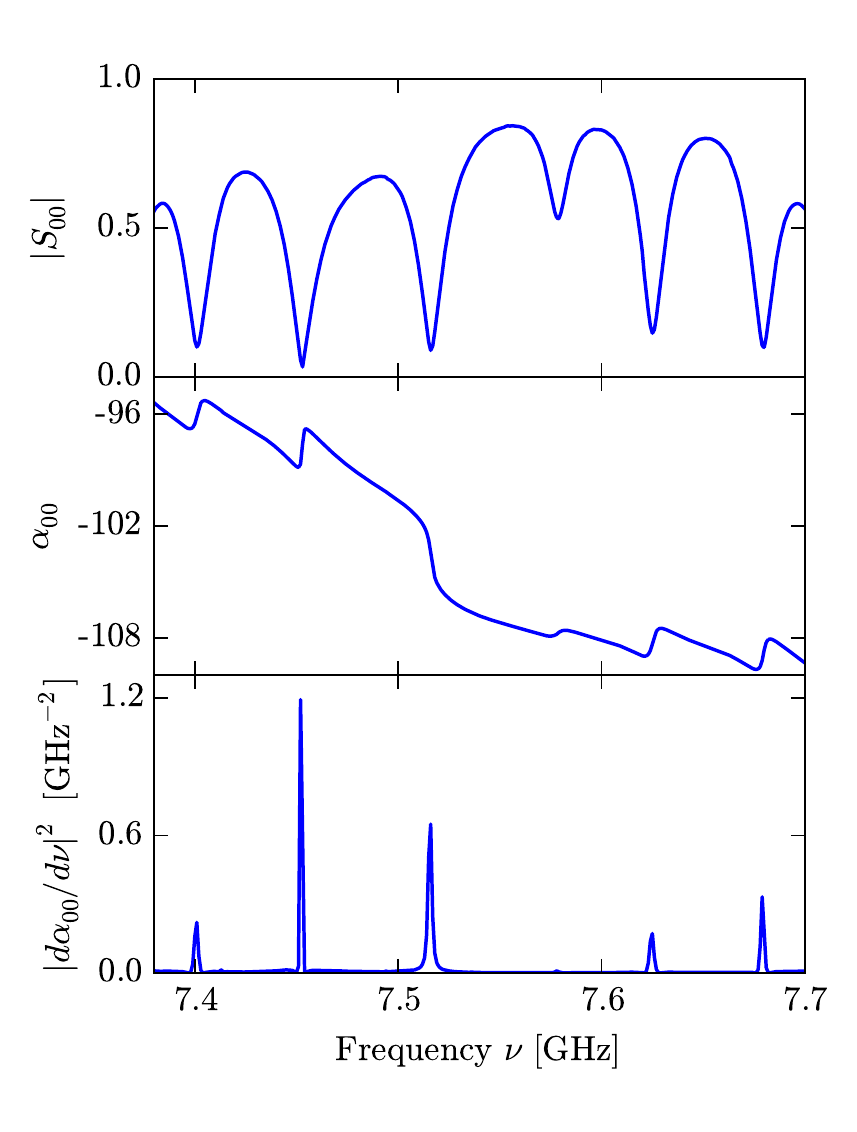}
  \caption{\label{fig:phasejumps} (color online)
  (a) Reflection $\left|S_{00}\right|^2$ for $\Delta\varphi=\pi$ for the same frequency range as in Fig.~\ref{fig:S11_spectra}.
  (b) Phase $\alpha_{00}$ of the reflection. (c) Squared phase derivative $\left|\mathrm{d}\alpha_{00}/\mathrm{d}\nu\right|^2$.
  }
\end{figure}
In the standard approach the eigenfrequencies are obtained by fitting Lorentzian lines to each resonance. But in the present situation there is an alternative. Because of the absence of transmission each resonance shows up as a jump in the reflection phase, as was explained in section~\ref{sec:scatt}. This is illustrated in Fig.~\ref{fig:phasejumps}. The upper panel shows part of a reflection spectrum with the same six resonances as in Fig.~\ref{fig:S11_spectra}. The corresponding reflection phase is plotted in Fig.~\ref{fig:phasejumps}(b). It shows a jump for each resonance. According to Eq.~(\ref{eq:G00}) these jumps should be infinitely sharp but because of absorption in the bonds they acquire a finite width. The overall decrease of the phase with $k$ results from an imperfection in the calibration: The raw data contain phases resulting from the propagation of the microwaves through the cables connecting the graph to the networks analyzer. The calibration should remove these phases, but phase contributions from the connectors can hardly be avoided. Figure~\ref{fig:phasejumps}(c) finally shows the square modulus of the phase, where the contribution from the overall decrease had been subtracted before. All resonances now show up as narrow peaks, much narrower as the resonances in the original reflection spectrum in Fig.~\ref{fig:phasejumps}(a). Taking all peaks exceeding a predefined discriminator level, about 90\,\% of the eigenvalues could be determined automatically. With the additional information from the spectral level dynamics, see Fig.~\ref{fig:S11_spectra}, some more could be identified. For about 10\,\% percent of the Kramers doublets the double peak was split into two resonances due to experimental imperfections. Whenever it was evident from the level dynamics that the two resonances correspond to a split Kramers doublet the two resonances were replaced by a single one in between.
\begin{figure}
  \mbox{\raisebox{5.3cm}[0cm][0cm]{a)\hspace{-.3cm}}\includegraphics[width=\columnwidth]{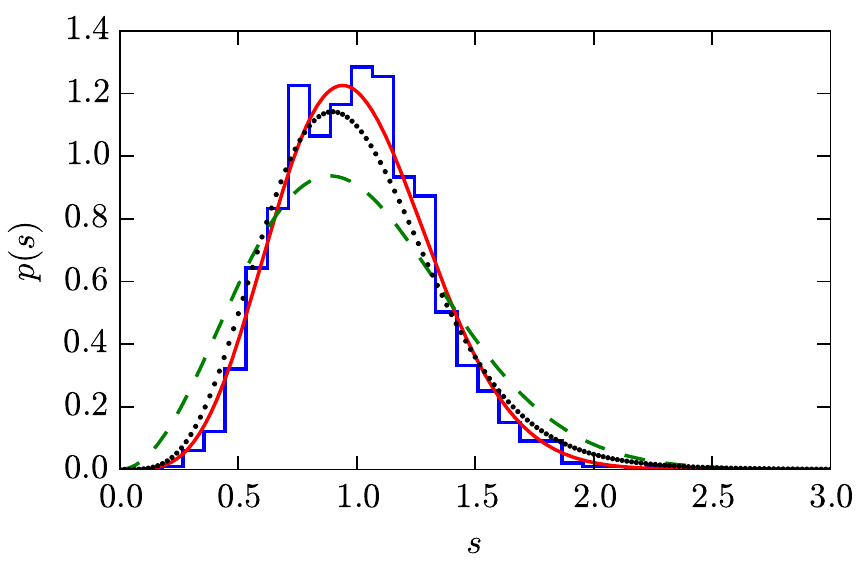}}
  \mbox{\raisebox{5.3cm}[0cm][0cm]{b)\hspace{-.3cm}}\includegraphics[width=\columnwidth]{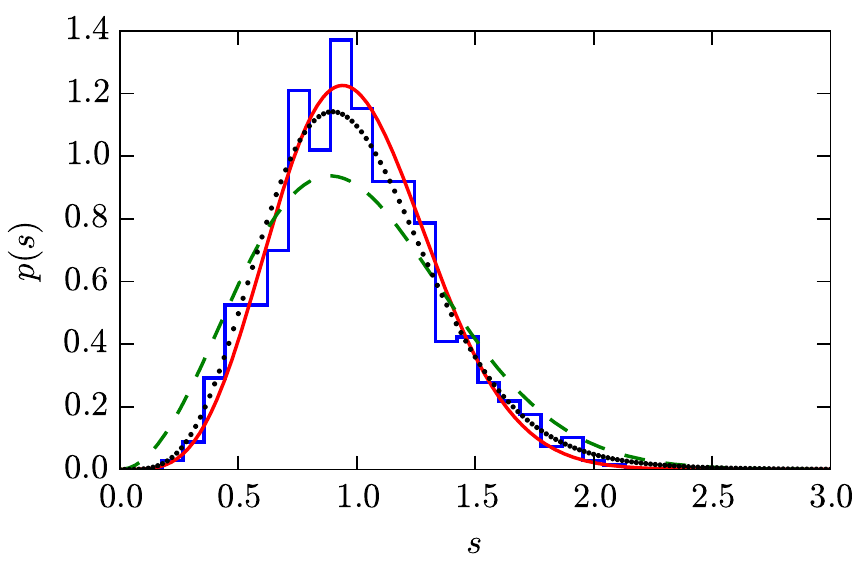}}
  \caption{\label{fig:wigner} (color online)
  (a) Spectral nearest neighbor spacing distribution for symplectic graphs with a single pair of connecting bonds obtained by superimposing the results from 9 different spectra (blue). The solid red and dashed green line correspond to GSE and GUE Wigner distributions, respectively, see Eqs.~(\ref{eq:GSE}) and (\ref{eq:GUE}). The dotted black line corresponds to the single pair of bonds Weyl approximation~(\ref{eq:psres}).
  (b) Same, but for symplectic graphs with two pairs of bonds.
  }
\end{figure}

From the Weyl formula~(\ref{eq:weyl}) it follows for the mean integrated density of states $n_\mathrm{Weyl}(k)=(\pi/L)k$, where $L$ is the total optical length of the graph. This allows to estimate the number of expected resonances. $L$ was determined from $L=n L_g$, where $L_g$ is the geometrical length including all cables and T-junctions, and where $n=1.43$ is the index of refraction of the dielectric within the coaxial cables. The number of found resonances in the average was by 8.5\,\% smaller than the value expected from the Weyl formula. A fraction of some percent of missing levels is not unusual in microwave studies. Generally the loss may have two sources: (i) misidentification of nearby resonances by a single one, (ii) missing of a resonance whenever a nodal line is close to the coupling point to the VNA. For GSE spectra the first source is ruled out because of the strong level repulsion, but the loss due to nodal lines remains. In principle it is possible to reduce the loss by nodal lines by repeating the measurements with different positions of the coupling points to the VNA. But since the additional effort would be considerable, and since this point is not of central importance in the present context, we refrained from doing this.

In section~\ref{sec:twobond} it was shown that one pair of bonds is not sufficient for a complete cross-over from two identical GUE spectra for the decoupled graphs to one GSE spectrum for the coupled graphs. Therefore, we performed two sets of measurements, one for graphs with only a single pair of bonds, the other one for graphs with two pairs of bonds. The resulting level spacing distributions are shown in Figs.~\ref{fig:wigner}(a) and (b), respectively. To improve the statistics, we superimposed the results from nine different graphs for the first case, and from five different graphs for the second case, each obtained by varying the coupling points for the bonds connecting the two subgraphs. In both cases the experimental results fit well to the GSE distribution and are clearly at odds with a GUE distribution. The dotted black line in Fig.~\ref{fig:wigner} corresponds to the single pair of bonds approximation~(\ref{eq:psres}). The statistics is not sufficient to discriminate between the two distributions. One would have to increase the number of resonances by a factor of 10 to see the difference. Furthermore, there is no visible difference between the level spacing distributions for the single pair of bonds and the two pair of bonds.

\subsection{Spectral correlations}\label{sec:corr}

\begin{figure}
  \includegraphics[width=\columnwidth]{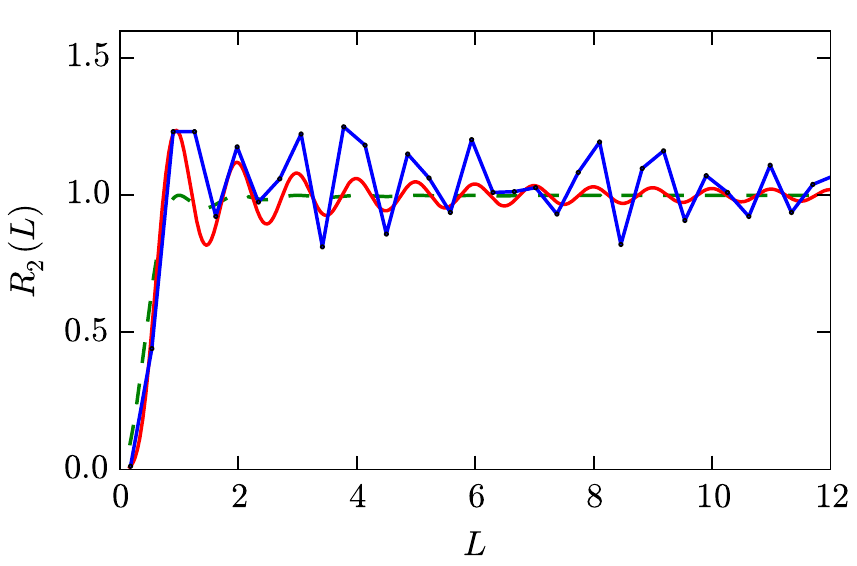}\\
  \caption{\label{fig:R2} (color online)
  Experimental two-point correlation function $R_2(L)$, together with the random matrix prediction for the GSE (solid red) and the GUE (dashed green), respectively.
  }
\end{figure}

\begin{figure}
  \includegraphics[width=\columnwidth]{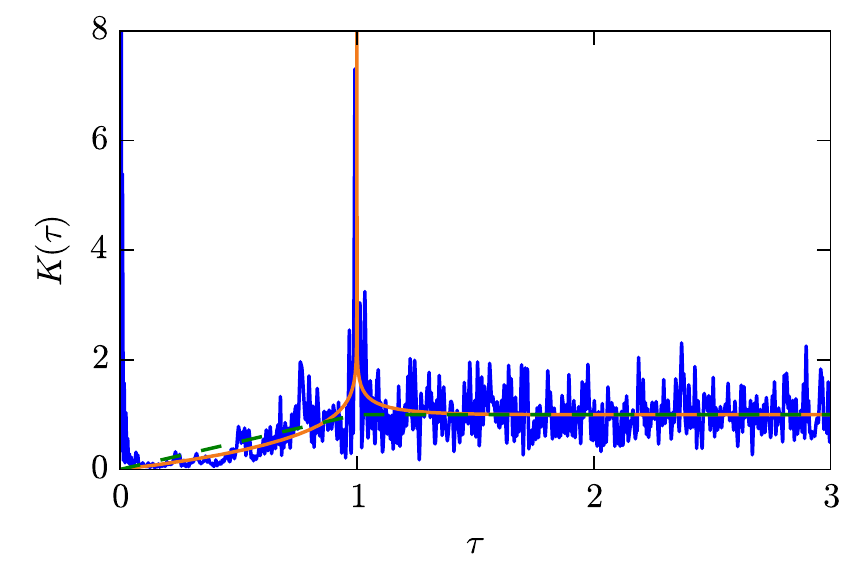}
  \caption{\label{fig:kgse} (color online)
  Experimental spectral form factor (blue) and the random matric expectations for the GSE (solid orange) and the GUE (dashed green), respectively.
  }
\end{figure}

One way to discriminate between the Gaussian ensembles is the level repulsion for short distances.
Another option is provided by the long distance behavior, in particular the spectral two-point correlation and quantities
derived from it. The spectral two-point function $R_2(L)$ is the probability density to find one
eigenvalue $E_N$ in the distance $L$ of another eigenvalue $E_M$,
\begin{equation}\label{eq:R2}
  R_2(L)=\frac{1}{N}\sum\limits_{n,m}{'}\,\delta\left(L-E_n+E_m\right)
\end{equation}
where $L$ is given in units of the mean level spacings. Random matrix results for the ensemble averaged two-point correlation functions for all three Gaussian ensembles can be found in literature \cite{meh91}. The two-point correlation function $R_2(L)$ shows oscillations for $L>1$ having their origin in the eigenvalue repulsion. The oscillations are present in all Gaussian ensembles, but are most prominent for the GSE because of the large repulsion exponent of 4. Figure~\ref{fig:R2}(a) shows the experimental and the theoretical $R_2(L)$ for the GSE graphs using the same data set as in Fig.~\ref{fig:wigner}. For comparison $R_2(L)$ for the GUE is plotted as well. The experimental data coincide well with the GSE prediction.

The most conspicuous quantity to discriminate between the different Gaussian ensembles is the Fourier transform of the two-point correlation function, the spectral form factor
\begin{equation}\label{eq:k}
  K(\tau)=\frac{1}{N}\sum\limits_{n,m}{'}\, e^{2\pi \imath\tau (E_n-E_m)}\,.
\end{equation}
For the GSE the ensemble averaged form factor is given by
\begin{equation}\label{eq:kgse}
  K(\tau)=\left\{\begin{array}{cc}
    \frac{1}{2}|\tau|-\frac{1}{4}|\tau|\ln|1-|\tau|| &\tau\leq 2 \\
    1  &\tau>2
  \end{array}\right.
\end{equation}
Equation~(\ref{eq:kgse}) describes a logarithmic singularity at $\tau=1$. This is a unique feature of the GSE as for the
other Gaussian ensemble $K(\tau)$ is continuous everywhere. Figure~\ref{fig:kgse} shows the experimental $K(\tau)$, together with the random matrix predictions for the GSE and the GUE, respectively. The experimental result follows
closely the GSE prediction, including the logarithmic singularity at $\tau=1$. In the experimental $K(\tau)$ the diagonal term of the double sum in Eq.~(\ref{eq:k}) was not removed but kept resulting in an extra peak at $\tau=0$.
\begin{figure}
  \mbox{\raisebox{5.1cm}[0cm][0cm]{a)\hspace{-.2cm}}\includegraphics[width=\columnwidth]{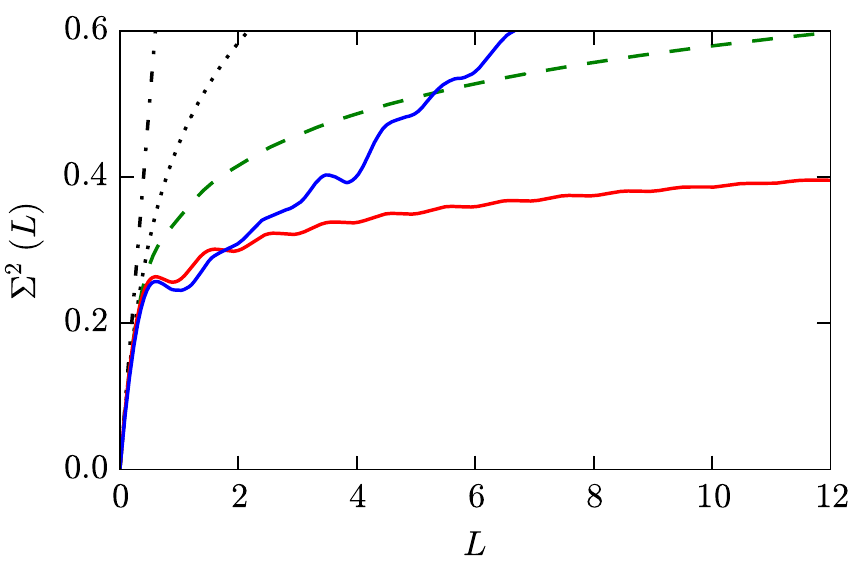}}
  \mbox{\raisebox{5.1cm}[0cm][0cm]{b)\hspace{-.2cm}}\includegraphics[width=\columnwidth]{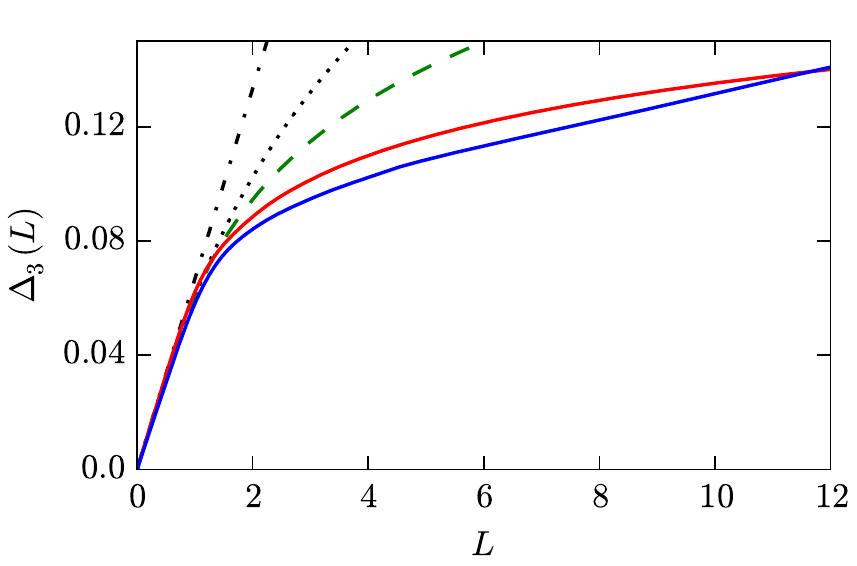}}
  \caption{\label{fig:sigma2} (color online)
  (a) Experimental number variance $\Sigma^2(L)$ and (b)~spectral rigidity $\Delta_3(L)$ using the same set of data as before.
  In addition the random matrix prediction for the Poisson ensemble (dashed dotted), the GOE (dotted), the GUE (dashed) and the GSE (solid) are shown.
  }
\end{figure}

In most cases the statistical evidence of the available data is not sufficient to study spectral two-point
correlation function and spectral form factor. Therefore, usually smoothed quantities are considered such as the number
variance $\Sigma^2(L)$, the variance of the number of levels in an interval of length $L$. Even more popular is the
spectral rigidity $\Delta_3(L)$. It is defined via the least-squares fit minimum of a fit of a linear line to the
integrated density of states $n(E)$ over an interval of length $L$. Alternatively, it may be written as
\begin{equation}\label{eq:Delta3}
\Delta_3(L)=\frac{2}{L^4}\int\limits_0^L \left(L^3-2L^2E+E^3\right)\Sigma^2(E)\mathrm{d}E\,.
\end{equation}
It may thus be interpreted as a smoothed version of $\Sigma^2(L)$ and hence a two-fold smoothed version of $R_2(E)$. Our
results for $\Sigma^2(L)$ and $\Delta_3(L)$ are plotted in Fig.~\ref{fig:sigma2}. In both cases there is a good agreement
with random matrix predictions for the GSE below $L=2$, but for larger $L$ values there are clear deviations which again
can be ascribed to missing levels. For a detailed discussion of the influence of missing levels on
$\Sigma^2(L)$ and $\Delta_3(L)$ we refer to Ref.~\onlinecite{bia16a}.

\subsection{From the GSE to the GOE}\label{sec:GSE_GOE}

\begin{figure*}
  \mbox{\raisebox{3.3cm}[0cm][0cm]{a)\hspace{-.2cm}}\includegraphics[width=0.666\columnwidth]{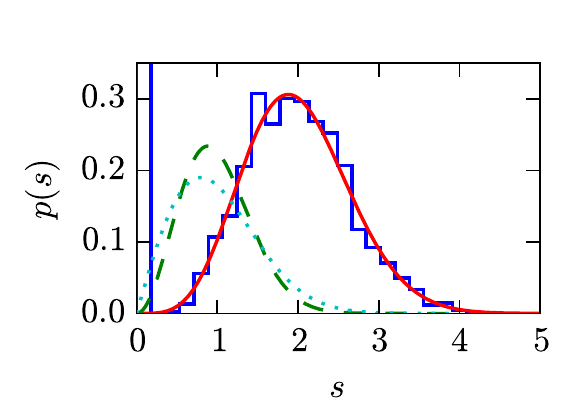}}
  \mbox{\raisebox{3.3cm}[0cm][0cm]{b)\hspace{-.2cm}}\includegraphics[width=0.666\columnwidth]{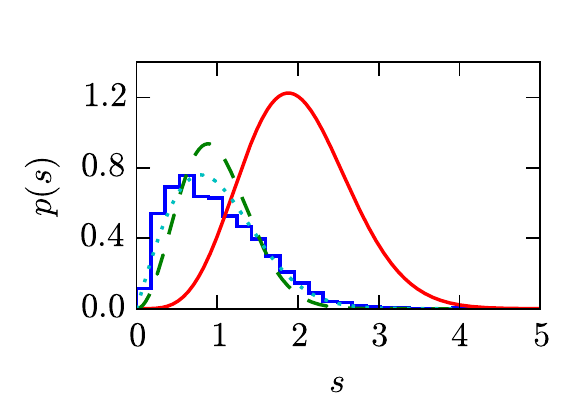}}
  \mbox{\raisebox{3.3cm}[0cm][0cm]{c)\hspace{-.2cm}}\includegraphics[width=0.666\columnwidth]{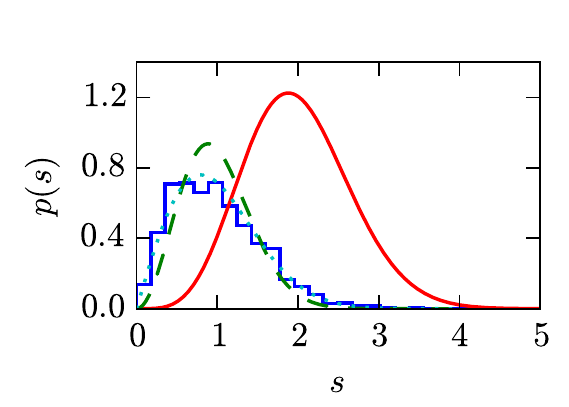}}\\
  \mbox{\raisebox{3.3cm}[0cm][0cm]{d)\hspace{-.2cm}}\includegraphics[width=0.666\columnwidth]{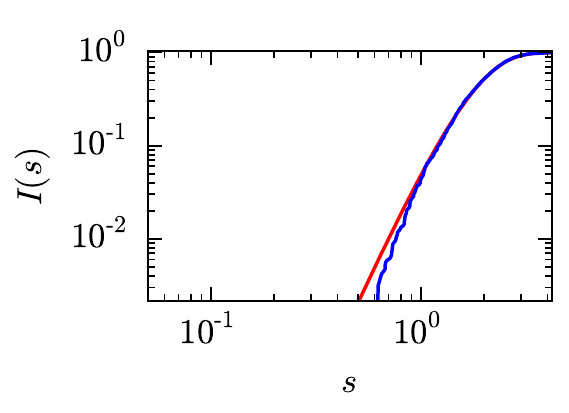}}
  \mbox{\raisebox{3.3cm}[0cm][0cm]{e)\hspace{-.2cm}}\includegraphics[width=0.666\columnwidth]{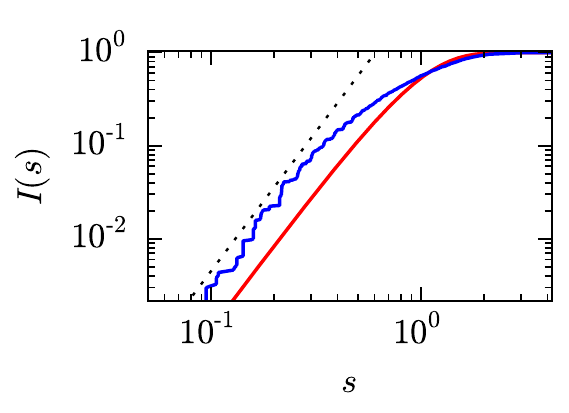}}
  \mbox{\raisebox{3.3cm}[0cm][0cm]{f)\hspace{-.2cm}}\includegraphics[width=0.666\columnwidth]{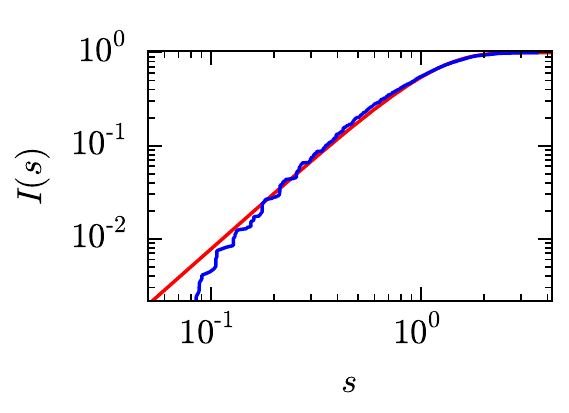}}\\
  \caption{\label{fig:gsegoe} (color online)
  Level spacing distributions [top, a)-c)] and corresponding integrated level spacing distributions [bottom, d)-f)] for $\Delta\varphi$=$\pi$, $\Delta\varphi$=1.5$\pi$ and $\Delta\varphi$=2$\pi$ (from left to right). The green dashed, gray dashed, and red solid lines correspond to the Wigner GOE, GUE, and GSE distributions, respectively.
  For the integrated spacings only the respective distribution is plotted as solid line. The dotted line in e) indicates a slope of 3.
  }
\end{figure*}

From the experiment the spectra are available for all $\Delta\varphi$ between $0$ and $2\pi$ which allows to study the spectral level dynamics as a function of $\Delta\varphi$. The situation $\Delta\varphi=\pi$ has been discussed already in detail before. It corresponds to the existence of an antiunitary symmetry $\mathcal{T}$ obeying $\mathcal{T}^2=-1$ resulting in Kramers degeneracy and GSE statistics for the Kramers doublets (with the proviso discussed in section~\ref{sec:twobond}).
For $\Delta\varphi=0$ or $\Delta\varphi=2\pi$ the minus signs in Eq.~(\ref{eq:sec3}) are missing. There is still an antiunitary symmetry $\mathcal{T}$, but now it squares to $+1$ corresponding to GOE statistics (see e.\,g. section 2.5 of Ref.~\onlinecite{haa01b}). For any $\Delta\varphi$ in between neither $\mathcal{T}^2=-1$ nor $\mathcal{T}^2=+1$ holds suggesting GUE statistics for this situation.

But there is another aspect to be considered. A change of $\Delta\varphi$ means a change of $\Delta l=\Delta\varphi/k$ and thus a change of the mean level density by $\Delta\rho_\mathrm{Weyl}=\Delta l/\pi=\Delta\varphi/(\pi k)$. Varying $\Delta\varphi$ from $0$ to $2\pi$ means thus a change of $\rho_\mathrm{Weyl}$ by $2/k$. The mean integrated density of states $n_\mathrm{Weyl}(k)=k\rho_\mathrm{Weyl}$ hence changes by 2, if $\varphi$ is changed by $2\pi$. In fact this statement is true not only for the mean, but also for the exact integrated density of states. This is a consequence of the periodicity of the spectra when changing $\varphi$ from $0$ to $2\pi$. This becomes obvious when looking into Fig.~\ref{fig:S11_spectra}. Following the resonance indicated by the white arrow for $\Delta\varphi=0$ and $\Delta\varphi=2\pi$ to the right it arrives two resonances later at the next crossing of the $2\pi$ line (blue arrow). Wrapping the spectra onto a cylinder by identifying the line $\Delta\varphi=0$ with the line $\Delta\varphi=2\pi$ the spectra may hence be interpreted in terms of just one pair of resonances twisted spiral-like along the cylinder surface. The two members of the pair merge whenever they cross the $\pi$ line and typically have a maximal spacing when crossing the $2\pi$ line.

For the moment we restricted the analysis to the spectra for $\Delta\varphi=\pi$, $\Delta\varphi=3\pi/2$, and $\Delta\varphi=2\pi$. According to the above discussion we expect GSE, GUE, and GOE statistics for these $\Delta\varphi$ values, respectively. Unfortunately the technique to obtain the eigenvalues from the derivative of the reflection phase works for $\Delta\varphi=\pi$ only. Therefore, we determined the spectra for $\Delta\varphi=3\pi/2$ and $\Delta\varphi=2\pi$ just from the minima in the reflection. We are aware of the deficiencies of this method, which is well-known to be unreliable in particular in the presence of overlapping resonances. Therefore, the present state of the data analysis should be considered as preliminary.

The results are presented in Fig.~\ref{fig:gsegoe}. The upper row shows, from left to right, the level spacings distribution $p(s)$ for $\Delta\varphi=\pi$, $\Delta\varphi=3\pi/2$, and $\Delta\varphi=2\pi$, respectively. For the sake of a coherent discussion we switched for $\Delta\varphi=\pi$ from the spacing distribution of Kramers doublets to that of the individual levels, yielding an additional peak at $s=0$ resulting from the Kramers degeneracy. Furthermore, because of the normalization of the mean level spacing to one, the Kramers doublets distribution moved by a factor of two to the right. The experimental distributions had been obtained by collecting the results from all graphs with a single pair of bonds and with two pair of bonds, respectively. Again the statistical evidence was not sufficient to discriminate between the two level spacings distributions.

For $\Delta\varphi=\pi$ and $\Delta\varphi=2\pi$ the experimental level spacing distributions are in accordance with the Wigner GSE and GOE distributions, respectively, but for $\Delta\varphi=3\pi/2$ there are clear deviations from the GUE distribution. The three distributions differ mainly in their repulsion exponents $\beta$ for small $s$ values. For a study of the small distance behavior the integrated level spacing distributions $I(s)=\int_0^s p(\bar{s})\,\mathrm{d}\bar{s}$ is useful. Because of the integration we expect $I(s)\sim s^{\beta+1}$ for small distances, turning into a straight line with slope $\beta+1$ in a log-log plot. The lower part of Fig.~\ref{fig:gsegoe} shows the experimental results together with the theoretical expectation. For $\Delta\varphi=\pi$ and $\Delta\varphi=2\pi$ a good agreement is found over nearly two orders of magnitudes of $I(s)$. The antiunitary symmetry $\mathcal{T}$ with $\mathcal{T}^2=-1$ or $\mathcal{T}^2=+1$ thus really maps onto a spectral repulsion $p(s)\sim s^\beta$ with $\beta=4$ and $1$, respectively. For $\Delta\varphi=3\pi/2$ the found level spacing distribution is clearly at odds with a Wigner GUE distribution, but still shows the expected level repulsion with $\beta=2$ [indicated by the dotted line Fig.~\ref{fig:gsegoe}e)].

The level dynamics in dependence of $\Delta\varphi=\pi$ is not sufficient to destroy the correlation between the members of the Kramers doublets as it seems: When increasing $\Delta\varphi$ starting with $\Delta\varphi=\pi$, the Kramers doublets split but there still remains a correlation between the previous members of the Kramers doublets over some $\Delta\varphi$ range. The excess in the experimental $p(s)$ beyond the GUE prediction for $s<0.5$ in the central figure of Fig.~\ref{fig:gsegoe} could be an indication of this remaining correlation.

\section{Conclusions}\label{sec:conclusions}

In our recent paper \cite{reh16} we presented first results on spectra in a microwave graph with symplectic symmetry.
The Kramers doublets expected for such a system had been clearly observed, and their level spacing distribution could be well described by the GSE Wigner distribution. In the present paper more details on the theoretical background and details of the data analysis are given. In addition results on spectral two-point correlation and spectral form factor as well as number variance an spectral rigidity are given. Apart from deviations due to about 8.5\,\% of missing levels a good agreement with random matrix predictions for the GSE was found.

Since it is not obvious whether a single pair of bond is sufficient for a complete cross-over from two identical GUE spectra for the decoupled graphs to one GSE spectrum for the coupled graphs, a Wigner-like formula was developed based on a $4\times 4$ matrix with $2\times 2$ GUE sub-matrices on the diagonal block describing the decoupled graphs, and off-diagonal blocks describing the coupling. The resulting level spacing distribution shares the $s^4$ repulsion for small distances with the GSE distribution, but there are relative deviations between the two distributions of some percent, comparable to the deviations between the Wigner and the exact level spacing distributions \cite{haa01b}. The experimental evidence was not sufficient to resolve this difference, but in a random matrix simulation it could be verified.

Finally, the level dynamics in dependence of the phase $\Delta\varphi$ between the two coupling bonds was studied. In dependence of $\Delta\varphi$ one expects a transition from the GSE for $\Delta\varphi=\pi$ to the GOE for $\Delta\varphi=2\pi$, which was really found. In particular the different repulsion behavior of $p(s)\sim s^4$ for $\Delta\varphi=\pi$ and $p(s)\sim s^1$ for $\Delta\varphi=2 \pi$ could be verified. In between one would expect GUE behavior, which was found only for small spacings, however. Up to now, only the spectra for $\Delta\varphi=\pi$, $\Delta\varphi=2\pi$, and $\Delta\varphi=3\pi/2$ have been analyzed. For a better understanding of the features of the spectral level dynamics an analysis of all data is indispensable. From the theoretical side an extension of the single pair of bonds approximation to arbitrary values of $\Delta\varphi$ would be desirable but is not easy since for $\Delta\varphi \ne \pi$ the symplectic symmetry is destroyed.

\appendix
\section*{Appendix: Derivation of Eq.~(\ref{eq:pv})}\label{app:A}
We are looking for
\begin{equation}\label{a01}
  p_V(z)=\langle\delta(z-4\left|\tilde{V}_{12}\right|^2)\rangle\,,
\end{equation}
the distribution function of $4|\tilde{V}_{12}|^2$, where $\tilde{V}_{12}$ is given by Eq.~(\ref{eq:ham1}),
\begin{equation}\label{a00}
  \tilde{V}_{12}=\psi_{11}\psi_{22}-\psi_{12}\psi_{21}\,,
\end{equation}
and the brackets denote the average over the $\psi_{ij}$. Using the Fourier representation of the delta function, this may
be rewritten as
\begin{equation}\label{a02}
  p_V(z)=\frac{1}{2\pi} \int\limits_{-\infty}^\infty \mathrm{d}x e^{\imath tz}\hat{p}_V(4t)
\end{equation}
where
\begin{equation}\label{a03}
  \hat{p}_V(a)=\langle e^{-\imath a|\tilde{V}_{12}|^2}\rangle=\langle e^{-\imath a(\tilde{V}_{12R}^2+\tilde{V}_{12I}^2)}\rangle\,.
\end{equation}
The lower indices $R$ and $I$ denote real and imaginary part, respectively. Applying twice a Gauss-Fresnel
transformation,
\begin{equation}\label{a04}
  e^{-\imath a v^2}=\sqrt{\frac{a}{\imath\pi}}\int\limits_{-\infty}^\infty e^{\imath a(x^2+2xv)}\mathrm{d}x
\end{equation}
one obtains
\begin{equation}\label{a05}
  \hat{p}_V(a)=\frac{a}{\imath \pi}\int\limits_{-\infty}^\infty \mathrm{d}x \int\limits_{-\infty}^\infty \mathrm{d}y e^{\imath a(x^2+y^2)}\,S
\end{equation}
with
\begin{equation}\label{a06}
  S= \left< e^{2\imath a[ x(\tilde{V}_{12})_R +y(\tilde{V}_{12})_I]}\right>\,.
\end{equation}
Using the assumption that all $\psi_{ij}$ are uncorrelated, the average factorizes,
\begin{equation}\label{a07}
  S=TT^*
\end{equation}
where
\begin{equation}\label{a08a}
  T=\left< e^{2\imath a[ x(\psi_{11}\psi_{22})_R +y(\psi_{11}\psi_{22})_I]}\right>\,.
\end{equation}
With the same distribution function~(\ref{eq:norm_ps}) for all $\psi_{ij}$ one obtains for the average
\begin{eqnarray}\label{a08b}
  T &=& \int\hspace{-0.5em}\int\hspace{-0.5em}\int\hspace{-0.5em}\int \mathrm{d}(\psi_{11})_R\,\mathrm{d}(\psi_{11})_I\,\mathrm{d}(\psi_{22})_R\,\mathrm{d}(\psi_{22})_I\\\nonumber
    &&\qquad\times e^{-\pi[(\psi_{11})_R^2+(\psi_{11})_I^2+(\psi_{22})_R^2+(\psi_{22})_I^2]}\\\nonumber
    &&\qquad\times e^{2\imath a\{x[(\psi_{11})_R(\psi_{22})_R-(\psi_{11})_I(\psi_{22})_I]\}}\\\nonumber
    &&\qquad\times e^{2\imath a\{y[(\psi_{11})_R(\psi_{22})_I+(\psi_{11})_I(\psi_{22})_R]\}}\\\nonumber
    &=&\int \mathrm{d}\psi\, e^{-\pi\Psi^TM\Psi}
\end{eqnarray}
where in the last step we introduced a short-hand notation with
\begin{equation}\label{a09}
    \Psi=\left(
           \begin{array}{c}
             (\psi_{11})_R \\
             (\psi_{11})_I \\
             (\psi_{22})_R \\
             (\psi_{22})_I\\
           \end{array}
         \right)\,,\quad
  M=\left(
      \begin{array}{cccc}
        1 & \cdot & \lambda x & \lambda y \\
        \cdot & 1 & \lambda y & -\lambda x \\
        \lambda x & \lambda y & 1 & \cdot \\
        \lambda y & -\lambda x & \cdot & 1 \\
      \end{array}
    \right)
\end{equation}
with $\lambda=-iaq/\pi$. The multidimensional Gauss integration of~(\ref{a08b}) yields
\begin{equation}\label{a10}
  T=\frac{1}{\sqrt{|M|}}= \left[1+\frac{a^2}{\pi^2}\left(x^2+y^2\right)\right]^{-1}\,.
\end{equation}
Plugging in this result into Eq.~(\ref{a07}), and substituting the variables
\begin{equation}\label{a11}
  x=\sqrt{\frac{\imath}{a}}r\cos\varphi\,\quad y=\sqrt{\frac{\imath}{a}}r\sin\varphi
\end{equation}
Eq.~(\ref{a05}) yields
\begin{equation}\label{a12}
  \hat{p}_V(a)=2\int\limits_0^\infty \mathrm{d}r\,r e^{-r^2}\left(1+\frac{\imath a}{\pi^2}r^2\right)^{-2}\,.
\end{equation}
Inserting this result into Eq.~(\ref{a01}) and changing the order of integrations one has
\begin{equation}\label{a13}
  \hat{p}_V(a)=\frac{1}{\pi}\int\limits_0^\infty \mathrm{d}r\,r e^{-r^2}f(r,z)
\end{equation}
where
\begin{equation}\label{a14}
  f(r,z)=\int_{-\infty}^\infty \mathrm{d}t e^{\imath tz}\left(1+\frac{4\imath t}{\pi^2}r^2\right)^{-2}\,.
\end{equation}
The latter integral can be solved by means of the residuum technique,
\begin{equation}\label{a15}
  f(r,z)=2\pi\left(\frac{\pi^2}{4r^2}\right)^2 z e^{-\frac{\pi^2}{4r^2}z}
\end{equation}
The remaining $r$ integral~(\ref{a13}) can be expressed in terms of an integral representation of $K_1(z)$ resulting in
Eq.~(\ref{eq:pv}).

\begin{acknowledgments}
This work was funded by the Deutsche Forschungsgemeinschaft via the individual grants STO 157/16-1 and KU 1525/3-1
and the European Commission through the H2020 programme by the Open Future Emerging Technology ``NEMF21'' Project (664828).
\end{acknowledgments}

\end{document}